# >3 W/mm Output Power $Al_{0.61}Ga_{0.39}N$ Channel HEMT with 24.4% associated PAE at 10 GHz

Meherab Hossain[1]*, Ruixin Bai[1], Swarnav Mukhopadhyay[1], Brahmani Challa[1], Kuan Chang Pan[2], Ryan Gilbert[2], Guangying Wang[1], Surjava Sanyal[1], Chuang Ye[1], Md Tahmidul Alam[1], Khush Gohel[1], Ahmad E. Islam[3], Andrew J. Green[3], Shubhra S. Pasayat[1], and Chirag Gupta[1]

[1]*Department of Electrical and Computer Engineering, University of Wisconsin-Madison, WI 53706, USA*

[2]*KBR, Wright-Patterson AFB, Dayton, OH 45433, USA*

[3]*Air Force Research Laboratory, Sensors Directorate, Wright-Patterson AFB, Dayton, OH 45433, USA*

*E-mail: meherab.hossain@wisc.edu

In this work, an ultra-wide bandgap (UWBG) high-electron-mobility transistor (HEMT) was fabricated using an $Al_{0.76}Ga_{0.24}N$ barrier and an $Al_{0.61}Ga_{0.39}N$ channel with thick $SiO_2$ passivation. The device demonstrated low current collapse at 40 V drain bias. At 10 GHz, the device exhibited an output power of 3.07 W/mm with an associated power-added efficiency (PAE) of 24.4%. A peak PAE of 33.9% was achieved at an associated output power of 2.14 W/mm. This excellent large-signal performance highlights the strong potential of UWBG AlGaN channel devices for X-band applications.

Ultra-wide bandgap (UWBG) materials promise high critical electric fields [1], [2], [3] and consequently higher power-handling capability for RF devices. While different ultra-wide bandgap materials have been explored for high-power [4], [5], [6], [7], [8], [9] and high-frequency applications [3], [10], [11], [12], [13], [14], [15], [16], [17], [18], [19], [20], [21], [22], [23], [24], [25], $Al_xGa_{1-x}N$ (x > 0.6) channel high-electron-mobility transistors (HEMTs) have consistently shown simultaneous high breakdown voltage and high-frequency performance [11], [12], [13], [16]. Johnson's figure of merit (JFOM), which represents the trade-off between high-frequency performance and high breakdown voltage, has surpassed 20 THz·V, with a cutoff frequency ($f_T$) of 22.5 GHz and a breakdown voltage of 930 V [12]. Additionally, a cutoff frequency of 27 GHz with a maximum oscillation frequency ($f_{max}$) of 70 GHz has been demonstrated for an AlGaN HEMT [11]. Several works on UWBG AlGaN channel polarization-graded field-effect transistors (PolFETs) have reported high frequency performance, where $f_T/f_{max}$ of 85 GHz/20 GHz [26] and 24.1 GHz/78.7 GHz [27] were obtained, although the breakdown performance of these devices was limited.

Although both small-signal frequency response and breakdown voltage are important indicators of device potential, large-signal load-pull characterization is particularly important to evaluate the device performance for practical applications, as it shows actual output power ($P_{out}$) and power-added efficiency (PAE) of a device at target frequencies. In UWBG materials, at X-band (8-12 GHz), a high output power of 2.7 W/mm (with ~2% PAE) [28] and a high power-added efficiency of ~22% (with output power of 1.8 W/mm) [15] have been demonstrated with AlGaN channel HEMTs. Recently, our team has reported 2.3 W/mm $P_{out}$ and 22.86% associated PAE at 20 GHz with AlGaN channel HEMTs [11]. So far, there are only a handful of reports on the large-signal operation of UWBG materials at X-band or above [11], [15], [28], [29], [30].

To achieve high output power and high power-added efficiency, it is important to have low on-resistance ($R_{on}$), which consists of contact resistance ($R_c$) and sheet resistance ($R_{sh}$). In an AlGaN channel HEMT epitaxial structure with a reverse-graded layer, one of the prominent factors affecting $R_c$ is the conduction band offset between the barrier and the channel [2]. The contact resistance can be reduced with a lower Aluminum compositional difference between the barrier and the channel. However, lowering the compositional difference also increases the $R_{sh}$ [3], [17]. Previously reported $Al_xGa_{1-x}N$ (x>60%) channel RF HEMTs (where at least small-signal performance has been reported) with ~20% compositional difference demonstrated sheet resistance between 3 kΩ/□ and 4 kΩ/□, with a $R_c$ of ≥3.5 Ω·mm [11], [12], [13], [18]. While a low $R_c$ of 1.55 Ω·mm was reported for UWBG AlGaN channel RF HEMTs, using a 9% compositional difference and an additional heavily doped thin barrier layer between the main barrier and the channel, still the sheet resistance was relatively high at 4.79 kΩ/□ [17]. Therefore, there is a design trade-off in HEMTs between sheet resistance and contact resistance that needs to be studied. Recently, regrown ohmic contacts have been explored for AlGaN channel HEMTs [2], [31], which can potentially break this trade-off. Additionally, large-signal performance also needs to be explored in more detail in UWBG AlGaN channel RF HEMTs.

In this work, we have explored the performance of an RF HEMT with a 15% compositional difference between the barrier and the channel in a reverse graded contact layer structure to optimize the contact resistance and sheet resistance while studying its impact on the large-signal performance of UWBG AlGaN channel HEMTs. With this design, a low $R_c$ of 2.54 Ω·mm with a reasonable $R_{sh}$ of 3.48 kΩ/□ was obtained. Large-signal pulsed load pull measurements on these HEMTs demonstrated excellent results with a peak $P_{out}$ of 3.07 W/mm (with associated PAE of 24.4%) and a peak PAE of 33.9% (with associated power of 2.14 W/mm) at 10 GHz. To our best knowledge, both peak $P_{out}$ and peak PAE obtained in this work are record performance for any UWBG materials at X-band or above.

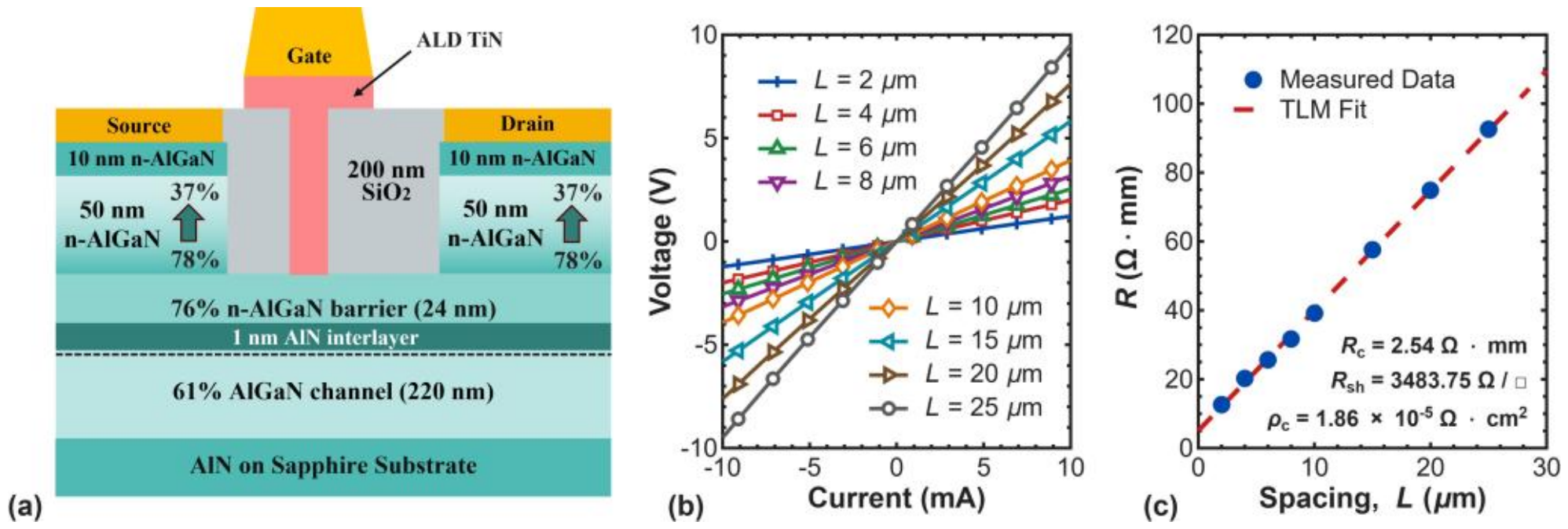

**Figure 1.** (a) Schematic cross-section of the device. (b) I-V curves for different TLM spacing. (c) Linear fit of the TLM data for $R_c$ and $R_{sh}$ extraction.

The device structure used for this work is shown in Fig. 1(a). The epitaxial structure was grown on sapphire using MOCVD. It contains a 220 nm $Al_{0.61}Ga_{0.39}N$ channel, a 24 nm n-doped $Al_{0.76}Ga_{0.24}N$ barrier, a 50 nm heavily n-doped reverse graded layer (Al composition 78% to 37%), and a 10 nm $Al_{0.37}Ga_{0.63}N$ heavily n-doped contact layer. Further details on the epitaxial growth can be found in the work by Mukhopadhyay et al. [32]. The small compositional difference between the barrier layer (~76%) and the start of the reverse-graded layer (~78%) has been introduced to further minimize contact resistance based on studies performed by Zhu et al. [33].

For device fabrication, a Ti/Al/Ni/Au (20/120/30/50 nm) metal stack was first deposited after optical lithography, followed by 30 seconds of annealing at 1000°C in nitrogen. A time-controlled low-power $Cl_2$-based reactive ion etch (RIE) was used to remove the reverse-graded layer, followed by mesa isolation. For passivation, a thick 200 nm $SiO_2$ layer was deposited using plasma-enhanced chemical vapor deposition (PECVD) at 350°C. Thereafter, electron beam lithography (EBL) was used for gate patterning, followed by $SiO_2$ hard mask removal using $CHF_3/CF_4/O_2$ plasma and atomic layer deposition (ALD) of TiN for Schottky gate. A metal stack of Cr/Au (35/365 nm) was used for the top gate after a second EBL patterning. Optical lithography and $CF_4/O_2$ plasma etch were used for via openings. Finally, Ti/Au (40/460 nm) was deposited for probe pads.

The device had a source-to-drain distance ($L_{sd}$) of 1.4 µm, a source-to-gate distance ($L_{sg}$) of 250 nm, a gate length ($L_g$) of 300 nm, and a total gate width ($W_g$) of 75 µm (2×37.5 µm). Due to lithographic challenges, lower gate length HEMTs were not resolved on this sample.

Fig. 1(b)-(c) shows the TLM data. A contact resistance ($R_c$) of 2.54 Ω·mm and a sheet resistance ($R_{sh}$) of 3483.75 Ω/□ were extracted from the TLM data as shown in Fig. 1(c). The $R_c$ obtained here is one of the lowest values reported for $Al_xGa_{1-x}N$ (x>60%) channel RF HEMTs [12], [13], [11], [18], [17]. This low $R_c$ can be attributed to the lower conduction band offset difference between the barrier and the channel, as well as the optimized reverse graded layer and contact layer. Previously, a state-of-the-art contact resistance of 1.55 Ω·mm was reported in a UWBG AlGaN channel RF HEMT [17]. However, that was achieved with a 9% compositional difference and an additional heavily doped 3 nm thin barrier layer between the main barrier layer and the channel [17], as opposed to ~15% compositional difference and a simpler barrier layer in this work. It should be noted that although lower contact resistances have been demonstrated in other AlGaN channel HEMTs, these have yet to be translated in RF HEMTs [8], [32].

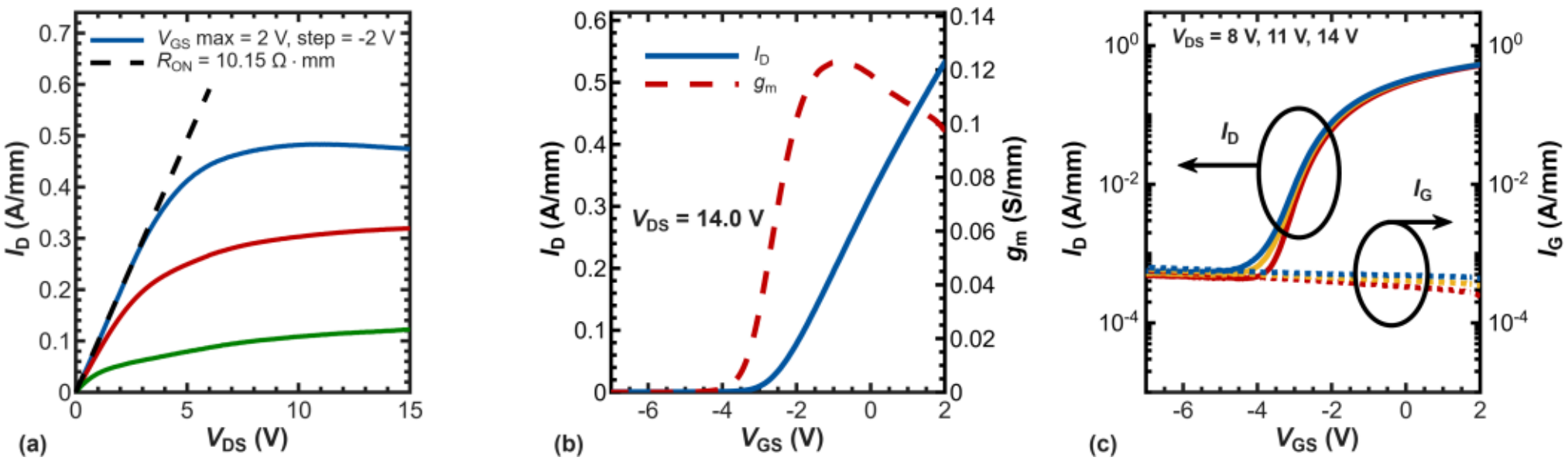

**Figure 2.** (a) DC output I-V characteristics of the device. (b) Transfer characteristics of the device on a linear scale. (c) Transfer characteristics of the device on a log scale at $V_{DS}$ = 8 V (red), 11 V (yellow), and 14 V (blue).

Fig. 2(a) shows the DC output characteristics of the device. A peak drain current ($I_D$) of 0.48 A/mm was obtained at $V_{GS}$ = 2 V and $V_{DS}$ = 11 V. The device exhibited good linear contact with an on-resistance ($R_{on}$) of 10.15 Ω·mm. The transfer characteristics of the device are shown in Fig. 2(b)-(c). A peak transconductance of 0.123 S/mm was obtained at $V_{GS}$ = -0.85 V and $V_{DS}$ = 14 V. This high transconductance (in spite of a relatively larger $L_g$) can be attributed to the low source resistance in these devices. A threshold voltage ($V_{th}$) of -3.92 V (defined at $I_{DS}$ = 1 mA/mm) was obtained at $V_{DS}$ = 14 V. Slightly higher gate and drain leakage currents have been observed in this device. The leakage may be caused by the passivation interface or etch damage during mesa isolation or reverse-grade etching. The identification of the exact mechanism and resolving this will be part of our future studies.

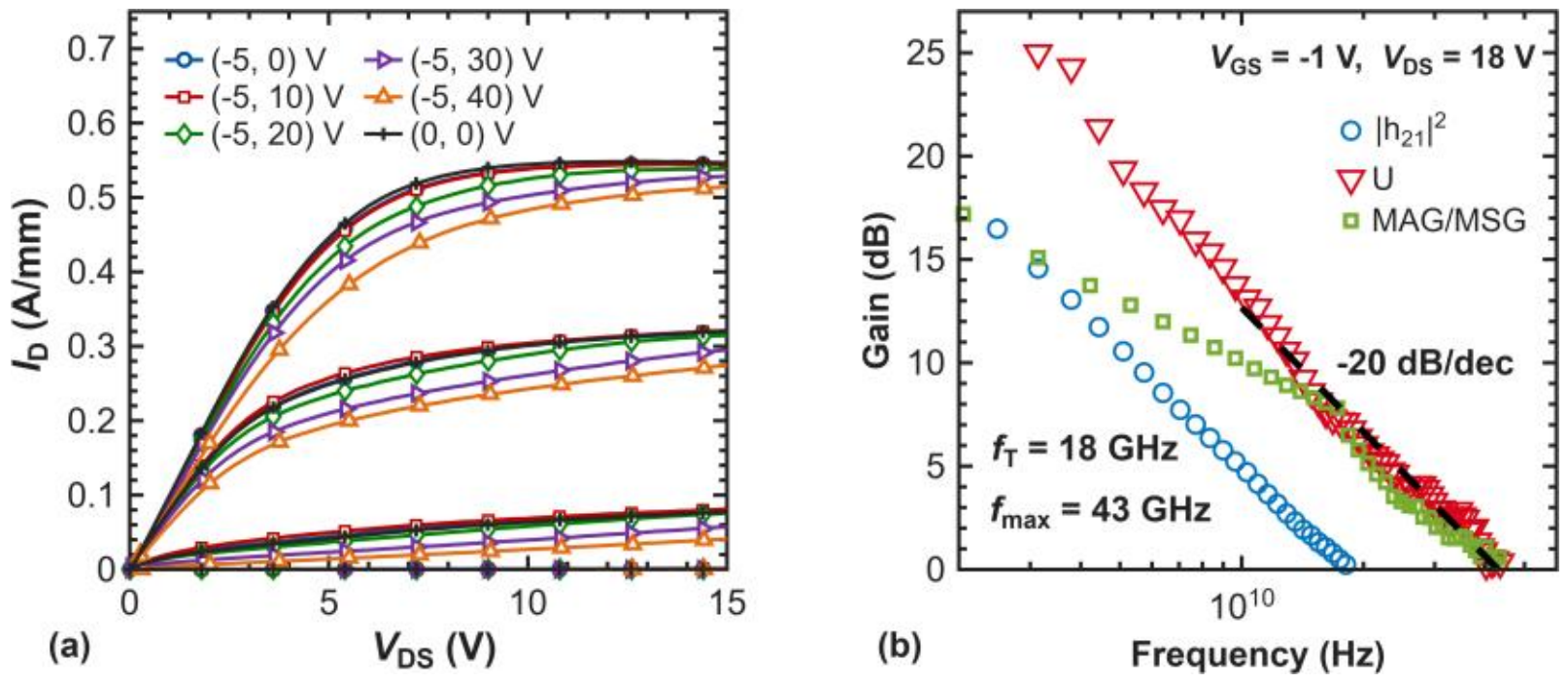

**Figure 3.** (a) Pulsed I-V characteristics of the device at zero quiescent bias and at different $V_{dsq}$ with $V_{gsq} < V_{th}$. (b) $|h_{21}|$, U, and MAG/MSG extracted from S-parameter measurements at ($V_{gsq}$, $V_{dsq}$) = (-1, 18) V.

Fig. 3(a) shows the pulsed IV characteristics of the device. The measurement was performed with a 50 µs pulse width and a 0.1% duty cycle. The pulse width was limited by the tool. The device showed good dispersion control, with low current collapse even at $V_{dsq}$ = 40 V and $V_{gsq} < V_{th}$. S-parameters for the device were measured with SOLT calibration and on-wafer open and short de-embedding. A cutoff frequency ($f_T$) of 18 GHz and a maximum oscillation frequency ($f_{max}$) of 43 GHz were extracted from the data as shown in Fig. 3(b). The relatively low small-signal frequency performance measured here, despite having high transconductance, is due to the large gate length of the device.

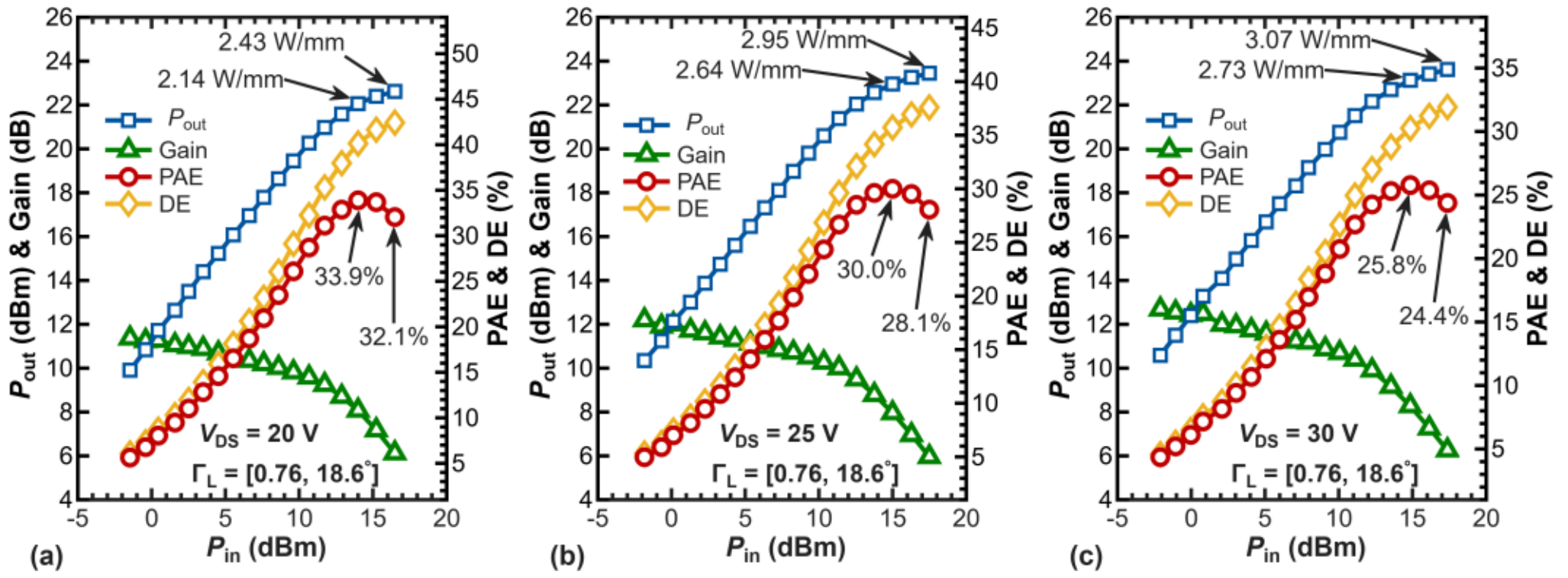

**Figure 4.** Pulsed 10 GHz class AB power sweep for the device with an $I_{dsq}$ of 100 mA/mm at (a) $V_{dsq}$ = 20 V, (b) $V_{dsq}$ = 25 V, and (c) $V_{dsq}$ = 30 V.

The large-signal performance of the device was evaluated with a passive pulsed load-pull setup at 10 GHz. A passive pulsed load pull was utilized to minimize the poor thermal conductivity effect of the sapphire substrate. For this measurement, a drain pulse width of 20 μs with a 1% duty cycle and an RF pulse width of 5 μs with a 0.25% duty cycle were used. The device was biased at class-AB with $I_{dsq}$ = 100 mA/mm. Fig. 4(a)-(c) shows the results from load-pull measurements with different $V_{dsq}$. An output power ($P_{out}$) of 3.07 W/mm with an associated power-added efficiency (PAE) of 24.4% was obtained for $V_{dsq}$ = 30 V. At Vdsq = 25 V, the device showed a PAE of 30% while the corresponding $P_{out}$ was 2.64 W/mm. Further reduction in drain voltage provided even more efficiency, with a maximum PAE reaching 33.9% and an associated $P_{out}$ of 2.14 W/mm at $V_{dsq}$ = 20 V. It should be noted that the large-signal performance reported in this work is limited by the maximum achievable load reflection coefficient ($\Gamma_L$) of the passive load-pull system. Since a passive tuner provides a limited range of achievable $\Gamma_L$, the optimal load-pull tuning point could not be fully reached, particularly at higher drain bias. In future studies, employing an active load-pull system or modifying the device design, such as using different gate widths to achieve a more favorable impedance range, could enable access to more optimal load-pull conditions. These approaches are expected to improve the large-signal performance and facilitate load-pull characterization at higher drain bias.

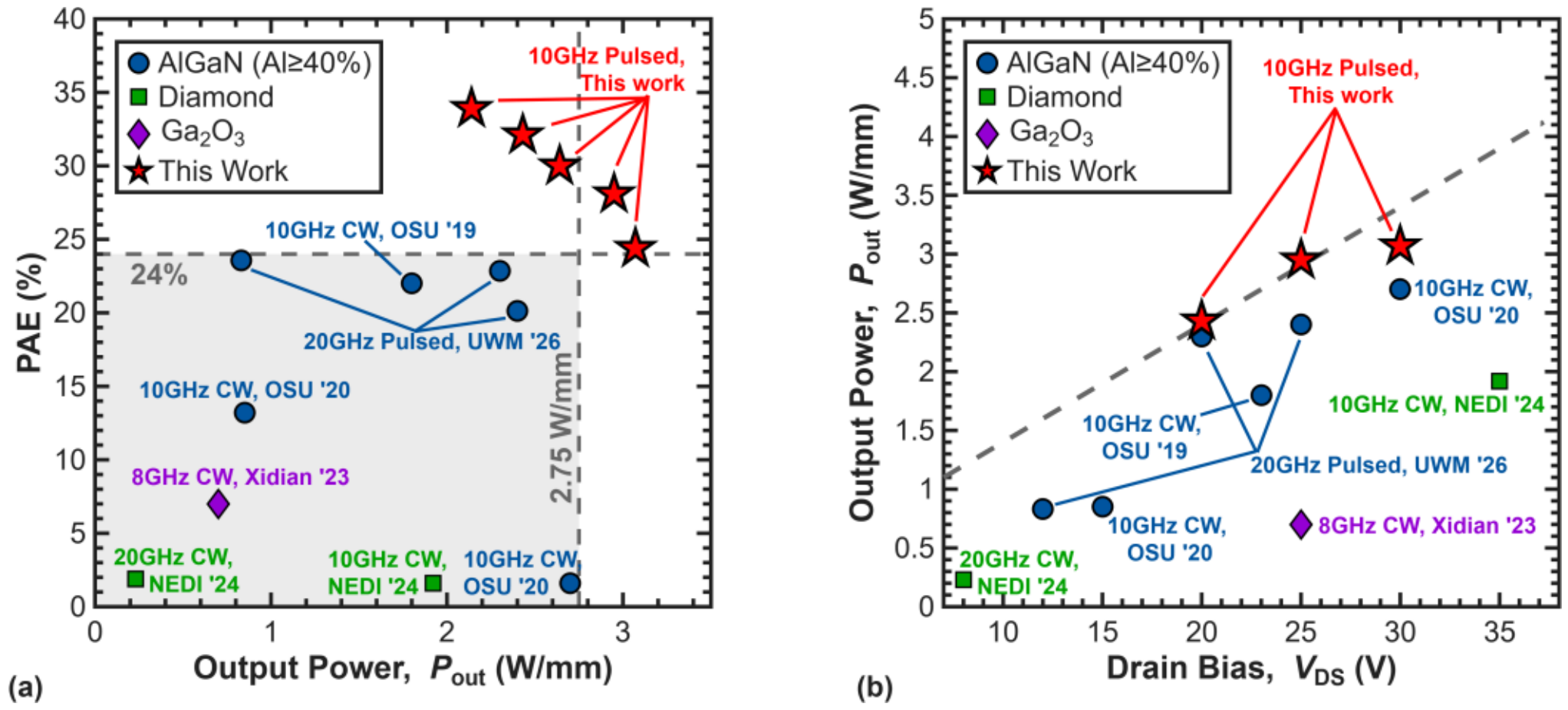

**Figure 5.** Benchmark for UWBG materials at X-band and above frequency range: (a) PAE vs. $P_{out}$, and (b) $P_{out}$ vs. $V_{dsq}$.

Fig. 5(a)-(b) shows large-signal performance benchmark for UWBG devices at X-band and above [11], [15], [28], [29], [30]. To the authors' best knowledge, the peak $P_{out}$ and peak PAE values achieved in this work are the highest values achieved for any UWBG material for the frequency range of ≥ 8 GHz [11], [15], [28], [29], [30]. The maximum $P_{out}$ of 3.07 W/mm obtained here is ~14% higher than the previously reported highest $P_{out}$ of 2.7 W/mm, while showing a more than ten-fold improvement in PAE at the same time (24.4% compared

to ~2%) [28]. Additionally, the maximum PAE of 33.9% achieved here is ~44% higher than the previously reported highest PAE of 23.58%, while having a simultaneous >150% improvement in $P_{out}$ (2.14 W/mm compared to 0.83 W/mm) [11].

In conclusion, we have explored a 15% Aluminum compositional difference between the barrier and channel for UWBG AlGaN channel RF HEMTs to improve large-signal performance. From our design with an $Al_{0.76}Ga_{0.24}N$ barrier and $Al_{0.61}Ga_{0.39}N$ channel, we obtained record results from load-pull measurements at 10 GHz. A peak $P_{out}$ of 3.07 W/mm and a peak PAE of 33.9% were demonstrated, both of which surpass previous records at X-band or above frequency ranges for any UWBG materials. These results highlight the tremendous potential of epitaxially engineered and well-passivated UWBG AlGaN HEMTs for next-generation high-frequency and high-power applications with excellent efficiency.

**Acknowledgments**

This work was supported in part by NSF ASCENT (award No. ECCS 2328137), ARPA-E Award DE- AR0001825, NSF Career Award ECCS 2441140, AFOSR Grant FA9550-23-10501, and FA9550-24-1-0159.

## References

[1] M. Farahmand, C. Garetto, E. Bellotti, K. F. Brennan, M. Goano, E. Ghillino, G. Ghione, J. D. Albrecht, and P. P. Ruden, “Monte Carlo simulation of electron transport in the III-nitride wurtzite phase materials system: binaries and ternaries,” *IEEE Transactions on Electron Devices*, vol. 48, no. 3, pp. 535–542, Mar. 2001, doi: 10.1109/16.906448.

[2] K. Ueno, R. Maeda, T. Kozaka, and H. Fujioka, “Degenerate GaN source–drain AlN/AlxGa1−xN/AlN high electron mobility transistors with a high breakdown electric field reaching 6.0 MV/cm,” *APL Mater.*, vol. 13, no. 4, p. 041129, Apr. 2025, doi: 10.1063/5.0249739.

[3] S. Shin, H. Pal, J. Pratt, J. Niroula, Y. Zhu, C. Joishi, B. A. Klein, A. Armstrong, A. A. Allerman, T. Palacios, and S. Rajan, “High breakdown electric field (>5 MV/cm) in UWBG AlGaN transistors,” *APL Electronic Devices*, vol. 1, no. 3, p. 036120, Aug. 2025, doi: 10.1063/5.0276718.

[4] A. D. Varma, S. Shin, J. McGlone, P. Darmawi-Iskandar, H. Pal, Z. Zhu, Y. Zhu, T. Jamil, S. M. T. Hasan, A. Al Mamun Mazumder, C. Joishi, J. Fiorenza, S. Rajan, A. Khan, and T. Palacios, “Ultra-Wide Bandgap Al0.85Ga0.15N/Al0.65Ga0.35N HEMTs with Baliga Figure of Merit > 0.5 GW/cm2 leveraging LPCVD SixNy Passivation and Field Plates,” in *2026 IEEE 38th International Symposium on Power Semiconductor Devices and ICs (ISPSD)*, May 2026, pp. 573–576. doi: 10.1109/ISPSD64561.2026.11553588.

[5] K. Gohel, B. Challa, S. Mukhopadhyay, G. Wang, M. T. Alam, R. Bai, J. Chen, S. Pasayat, and C. Gupta, “2.3 kV Al-rich AlGaN channel HEMTs on sapphire with >2.2 MV $cm^{-1}$ average e-field and >650 MW $cm^{-2}$ Baliga figure of merit,” *Appl. Phys. Express*, vol. 19, no. 6, p. 066502, Jun. 2026, doi: 10.35848/1882-0786/ae7531.

[6] M. T. Alam, B. Challa, S. Mukhopaddhay, G. Wang, K. Gohel, S. S. Pasayat, and C. Gupta, “2 kV monolithic bidirectional $Al_{0.61}$ $Ga_{0.39}$ N-channel HEMT-on-sapphire substrate,” *Jpn. J. Appl. Phys.*, vol. 65, no. 14, p. 140904, Jul. 2026, doi: 10.35848/1347-4065/ae7f61.

[7] K. Gohel, S. Mukhopadhyay, R. Imran Roya, S. Sanyal, M. T. Alam, J. Chen, R. Bai, G. Wang, S. Pasayat, and C. Gupta, “>2.7 kV Al0.65Ga0.35N Channel HEMT on Bulk AlN Substrate With >400 $MW/cm^2$ Baliga Figure of Merit,” *IEEE Electron Device Letters*, vol. 46, no. 11, pp. 2102–2105, Nov. 2025, doi: 10.1109/LED.2025.3606251.

[8] M. T. Alam, K. Gohel, B. Challa, S. Xie, S. Mukhopadhyay, G. Wang, S. Sanyal, R. Bai, S. S. Pasayat, and C. Gupta, “High-Voltage (600 V) Dynamic Switching in Al0.6Ga0.4N-Channel HEMTs With Multiple Field-Plates,” *IEEE Electron Device Letters*, vol. 47, no. 10, pp. 1999–2002, Oct. 2026, doi: 10.1109/LED.2026.3725547.

[9] S. Kosanovic, M. I. Khan, H. Rana, O. Odabasi, R. Baby, and E. Ahmadi, “First demonstration of an N-polar high Al-content AlGaN channel HEMT grown by plasma-assisted molecular beam epitaxy,” *APL Electronic Devices*, vol. 2, no. 1, p. 016128, Mar. 2026, doi: 10.1063/5.0300343.

[10] S. Shin, K. Liddy, J. Pratt, C. Cao, Y. Zhu, B. A. Klein, A. Armstrong, A. A. Allerman, and S. Rajan, “High breakdown field multi-kV UWBG AlGaN transistors,” *APL Electronic Devices*, vol. 2, no. 2, p. 026120, May 2026, doi: 10.1063/5.0337643.

[11] R. Bai, S. Mukhopadhyay, C. Desmier, K. Gohel, S. Sanyal, J. Chen, D. Matlock, E. Gebara, N. C. Miller, S. S. Pasayat, and C. Gupta, “K-Band (20 GHz) Large-Signal Performance in Al0.65Ga0.35N Channel HEMTs With 2.3-W/mm Output Power and 22.8% Associated PAE,” *IEEE Electron Device Letters*, vol. 47, no. 4, pp. 688–691, Apr. 2026, doi: 10.1109/LED.2026.3667057.

[12] R. Bai, S. Mukhopadhyay, K. Gohel, S. Sanyal, J. Chen, M. T. Alam, S. Xie, S. S. Pasayat, and C. Gupta, “Demonstration of high Johnson’s figure of merit ($f_t \times V_{BR}$ >20 THz·V) and $f_{max} \times V_{BR}$ (>42 THz·V) for $Al_{0.66}$ $Ga_{0.34}$ N channel MISHEMT on bulk AlN substrates,” *Appl. Phys. Express*, vol. 18, no. 8, p. 086501, Aug. 2025, doi: 10.35848/1882-0786/adf373.

[13] J. Chen, P. Seshadri, A. Al Mamun Mazumder, R. Bai, R. Rao, A. Khan, and C. Gupta, “High-Frequency (fT30 GHz) High Breakdown (>300 V) $Al_{0.62}Ga_{0.38}N$ Channel HEMT,” *IEEE Transactions on Electron Devices*, vol. 72, no. 9, pp. 4752–4756, Sep. 2025, doi: 10.1109/TED.2025.3584738.

[14] J. Chen, P. Seshadri, K. Stephenson, M. Abdullah Mamun, R. Bai, Z. Wang, A. Khan, and C. Gupta, “64% AlGaN Channel HFET With High Johnson’s Figure of Merit (>6 THz·V),” *IEEE Electron Device Letters*, vol. 46, no. 4, pp. 545–548, Apr. 2025, doi: 10.1109/LED.2025.3541145.

[15] H. Xue, C. H. Lee, K. Hussian, T. Razzak, M. Abdullah, Z. Xia, S. H. Sohel, A. Khan, S. Rajan, and W. Lu, “$Al_{0.75}$ $Ga_{0.25}$ N/$Al_{0.6}$ $Ga_{0.4}$ N heterojunction field effect transistor with $f_T$ of 40 GHz,” *Appl. Phys. Express*, vol. 12, no. 6, p. 066502, Jun. 2019, doi: 10.7567/1882-0786/ab1cf9.

[16] A. G. Baca, B. A. Klein, J. R. Wendt, S. M. Lepkowski, C. D. Nordquist, A. M. Armstrong, A. A. Allerman, E. A. Douglas, and R. J. Kaplar, “RF Performance of Al0.85Ga0.15N/Al0.70Ga0.30N High Electron Mobility Transistors With 80-nm Gates,” *IEEE Electron Device Letters*, vol. 40, no. 1, pp. 17–20, Jan. 2019, doi: 10.1109/LED.2018.2880429.

[17] S. Shin, C. Cao, J. Pratt, Y. Zhu, B. A. Klein, A. Armstrong, A. A. Allerman, and S. Rajan, “Barrier electrostatics and contact engineering for ultra-wide bandgap AlGaN HFETs,” *APL Electronic Devices*, vol. 1, no. 4, p. 046131, Dec. 2025, doi: 10.1063/5.0303729.

[18] J. Chen, A. Al Mamun Mazumder, P. Seshadri, D. Nandakumar, R. Bai, R. A. Choudhury, A. Khan, and C. Gupta, “Temperature dependent high frequency performance of a 62% AlGaN channel HEMT,” *APL Electronic Devices*, vol. 2, no. 1, p. 016102, Mar. 2026, doi: 10.1063/5.0301547.

[19] S. Imanishi, K. Horikawa, N. Oi, S. Okubo, T. Kageura, A. Hiraiwa, and H. Kawarada, “3.8 W/mm RF Power Density for ALD Al2O3-Based Two-Dimensional Hole Gas Diamond MOSFET Operating at Saturation Velocity,” *IEEE Electron Device Letters*, vol. 40, no. 2, pp. 279–282, Feb. 2019, doi: 10.1109/LED.2018.2886596.

[20] X. Yu, J. Zhou, C. Qi, Z. Cao, Y. Kong, and T. Chen, “A High Frequency Hydrogen-Terminated Diamond MISFET With fT/fmax of 70/80 GHz,” *IEEE Electron Device Letters*, vol. 39, no. 9, pp. 1373–1376, Sep. 2018, doi: 10.1109/LED.2018.2862158.

[21] C. N. Saha, A. Vaidya, A. F. M. A. U. Bhuiyan, L. Meng, S. Sharma, H. Zhao, and U. Singisetti, “Scaled β-Ga2O3 thin channel MOSFET with 5.4 MV/cm average breakdown field and near 50 GHz fMAX,” *Applied Physics Letters*, vol. 122, no. 18, p. 182106, May 2023, doi: 10.1063/5.0149062.

[22] X. Yu, W. Xu, Y. Wang, B. Qiao, R. Shen, J. Zhou, Z. Li, T. You, Z. Shen, K. Zhang, F.-F. Ren, D. Tang, X. Ou, G. Han, Y. Kong, T. Chen, S. Gu, Y. Zheng, J. Ye, and R. Zhang, “Heterointegrated Ga2O3-on-SiC RF MOSFETs With f T/f max of 47/51 GHz by Ion-Cutting Process,” *IEEE Electron Device Letters*, vol. 44, no. 12, pp. 1951–1954, Dec. 2023, doi: 10.1109/LED.2023.3327134.

[23] Z. Xia, H. Xue, C. Joishi, J. Mcglone, N. K. Kalarickal, S. H. Sohel, M. Brenner, A. Arehart, S. Ringel, S. Lodha, W. Lu, and S. Rajan, “β -Ga2O3 Delta-Doped Field-Effect Transistors With Current Gain Cutoff Frequency of 27 GHz,” *IEEE Electron Device Letters*, vol. 40, no. 7, pp. 1052–1055, Jul. 2019, doi: 10.1109/LED.2019.2920366.

[24] C. N. Saha, A. Vaidya, N. J. Nipu, L. Meng, D. S. Yu, H. Zhao, and U. Singisetti, "Thin channel Ga2O3 MOSFET with 55 GHz fMAX and > 100 V breakdown," *Applied Physics Letters*, vol. 125, no. 6, p. 062101, Aug. 2024, doi: 10.1063/5.0208580.

[25] H. Ye, M. Gaevski, G. Simin, A. Khan, and P. Fay, "Electron mobility and velocity in Al0.45Ga0.55N-channel ultra-wide bandgap HEMTs at high temperatures for RF power applications," *Applied Physics Letters*, vol. 120, no. 10, p. 103505, Mar. 2022, doi: 10.1063/5.0084022.

[26] Y. Zhu, A. A. Allerman, A. Wissel-Garcia, S. Shin, J. Pratt, C. Cao, K. J. Liddy, B. A. Klein, A. Armstrong, J. S. Speck, and S. Rajan, "Ultra-Wide Bandgap AlGaN Heterostructure Field Effect Transistors with Current Gain Cutoff Frequency Above 85 GHz," *IEEE Electron Device Letters*, pp. 1–1, 2026, doi: 10.1109/LED.2026.3733992.

[27] S. Kawasaki, M. Hiroki, K. Hirama, and Y. Taniyasu, "First RF Operation of AlGaN-channel Polarization-Doped FETs with Average Al-content over 0.75," in *2025 IEEE International Electron Devices Meeting (IEDM)*, Dec. 2025, pp. 1–4. doi: 10.1109/IEDM50572.2025.11353758.

[28] H. Xue, K. Hussain, T. Razzak, M. Gaevski, S. H. Sohel, S. Mollah, V. Talesara, A. Khan, S. Rajan, and W. Lu, "Al0.65Ga0.35N/Al0.4Ga0.6N Micro-Channel Heterojunction Field Effect Transistors With Current Density Over 900 mA/mm," *IEEE Electron Device Letters*, vol. 41, no. 5, pp. 677–680, May 2020, doi: 10.1109/LED.2020.2977997.

[29] X. Yu, B. Qiao, Y. Sun, R. Tao, R. Wang, J. Zhou, H. Guo, Z. Li, H. Lu, H. Gong, Z. Cao, Y. Zhang, J. Ye, Y. Kong, and T. Chen, "High-k/Diamond RF MOSFETs with Record High fT/fmax of 90/164 GHz and First Demonstration of a Diamond MMIC," in *2024 IEEE International Electron Devices Meeting (IEDM)*, Dec. 2024, pp. 1–4. doi: 10.1109/IEDM50854.2024.10873491.

[30] M. Zhou, H. Zhou, S. Huang, M. Si, Y. Zhang, T. Luan, H. Yue, K. Dang, C. Wang, Z. Liu, J. Zhang, and Y. Hao, "1.1 A/mm ß-Ga2O3-on-SiC RF MOSFETs with 2.3 W/mm Pout and 30% PAE at 2 GHz and fT/fmax of 27.6/57 GHz," in *2023 International Electron Devices Meeting (IEDM)*, Dec. 2023, pp. 1–4. doi: 10.1109/IEDM45741.2023.10413782.

[31] T. Jamil, A. A. M. Mazumder, S. M. T. Hasan, M. Rahman, M. Ali, A. Malik, K. Hussain, C. Joishi, M. Sadek, J. G. Fiorenza, G. Simin, and A. Khan, "Regrown n++-GaN ohmic contacts to high-Al content AlxGa1−xN (x > 0.6) digital alloy channels using AlN/GaN short-period superlattice," *Applied Physics Letters*, vol. 129, no. 4, p. 042108, Jul. 2026, doi: 10.1063/5.0338404.

[32] S. Mukhopadhyay, K. Gohel, S. Sanyal, M. Dangi, R. I. Roya, R. Bai, J. Chen, Q. Lin, G. Wang, C. Gupta, and S. S. Pasayat, "Characteristics of transport properties in ultra-wide bandgap Al0.65Ga0.35N channel HEMTs with low contact resistance and high breakdown voltage (>2.5 kV)," *Applied Physics Letters*, vol. 126, no. 15, p. 152103, Apr. 2025, doi: 10.1063/5.0256300.

[33] Y. Zhu, A. A. Allerman, C. Joishi, J. Pratt, A. M. Dominic Merwin Xavier, G. Calderon Ortiz, B. A. Klein, A. Armstrong, J. Hwang, and S. Rajan, "Heterostructure and interfacial engineering for low-resistance contacts to ultra-wide bandgap AlGaN," *Applied Physics Letters*, vol. 126, no. 6, p. 062107, Feb. 2025, doi: 10.1063/5.0230220.